\documentclass[twocolumn,preprintnumbers,amsmath,amssymb,superscriptaddress,nofootinbib,pre]{revtex4}
\usepackage{dcolumn}
\usepackage{bm}
\usepackage{graphics}
\usepackage{graphicx}
\usepackage{epstopdf}

\begin{document}

\title{Non-affine geometrization can lead to nonphysical instabilities}

\author{Eduardo Cuervo-Reyes}
\email{eduardo.cuervoreyes@empa.ch}
\affiliation{Swiss Federal Laboratories of Materials Science and Technology, \"{U}berlandstrasse 129, CH-8600 D{\"u}bendorf  Switzerland}

\affiliation{Swiss Federal Institute of Technology (ETH), CH-8093 Z\"{u}rich, Switzerland}

\author{Ramis Movassagh}
\email{ramis.mov@gmail.com}
\affiliation{Department of Mathematics, Northeastern University, Boston, MA, 02115}
 \affiliation{Department of Mathematics, Massachusetts Institute of Technology, Cambridge, MA, 02139}

\date{\today}
\begin{abstract}
Geometrization of dynamics consists of representing trajectories by  geodesics on a configuration space with a suitably defined metric. Previously, efforts were made to show that the analysis of dynamical stability can also be carried out within geometrical frameworks, by measuring the broadening rate of  a bundle of geodesics. Two known formalisms are via Jacobi and Eisenhart metrics. We find that this geometrical analysis measures the actual stability when the length of any geodesic is proportional to the corresponding time interval. We prove that the Jacobi metric is not always an appropriate parametrization by showing that it predicts chaotic behavior for a system of harmonic oscillators. Furthermore, we show, by explicit calculation, that the correspondence between dynamical- and geometrical-spread is ill-defined for the Jacobi metric. We find that the Eisenhart dynamics corresponds to the actual tangent dynamics and is therefore an appropriate geometrization scheme.  
\end{abstract}


\maketitle

\section{Stability of Hamiltonian Systems}
\subsection{Lyapunov exponent}
Many physical systems are well represented by the time evolution of the coordinates ($q^i_{}(t)$, $i=1\ldots N$) of particles with  inertia matrix $a^{}_{ij}(\bm{q})$, moving under the influence  of  the potential $V(\bm{q})$. The $q^i_{}(t)$, and the momenta $p_i^{}(t)=a^{}_{ij}\dot{q}^j_{}(t)$, where repeated indices are summed over, are found by integrating Hamilton's equations
\begin{equation}
\dot{p_{i}}=-\frac{\partial H}{\partial q_{i}}\; , \qquad \dot{q_{i}}=\frac{\partial H}{\partial p_{i}}\; ,\label{HEq}
\end{equation}
\noindent with the  Hamiltonian
\begin{equation}
H(\bm{q},\bm{p})=\frac{1}{2}a^{ij}(\bm{q})p_{i}p_{j}+V(\bm{q})\; , 
\end{equation}
\noindent where $a^{ik}_{}a^{}_{kj}=\delta^i_j$.  Since, in practical situations, initial conditions ($\bm{q}(t_0)=\bm{q}_0$, and $\bm{p}(t_0)=\bm{p}_0$) are subject to  uncertainties,  one often needs to know the extent to which a pair of trajectories, which are infinitesimally close at some time, will  remain close at later times; i.e., the extent to which the dynamics is stable. The stability of dynamical systems is also relevant in the context of statistical methods, as it has strong implications on the ergodicity of isolated systems.

A Hamiltonian system in general may have stable or unstable trajectories, depending on the region of the phase space where the dynamics takes place \cite{Poincare,Arnold,Krylov}. The degree of instability of a set of trajectories can be quantified by the Lyapunov exponent 
\begin{equation}
\lambda\equiv\lim_{t\rightarrow\infty}\frac{1}{t}\ln\left(\frac{||\bm{x}(t)||}{||\bm{x}(0)||}\right)\; ,
\end{equation} 
\noindent where $||\bm{x}(t)||$ is the norm of the $2N$-dimensional variation of the phase-space trajectory. Note that there are in principle $2N$ lyapunov exponents.  Since it is sufficient for detecting chaos, we will only consider the largest and simply refer to it as ``the'' lyapunov exponent. The variation $\bm{x}(t)$ is obtained from the solution of the linearized variation  of the equations of motion (the tangent dynamics) \cite{TD1,TD2,TD3,TD4}
\begin{subequations}\label{TD}
\begin{eqnarray}
\Delta {\dot p}^i=-\left(\Delta q^j\frac{\partial}{\partial q^j}+\Delta p_j\frac{\partial}{\partial p_j}\right)\frac{\partial H(\bm{q},\bm{p})}{\partial q^i}\; ,\\
\Delta {\dot q}^i=\left(\Delta q^j\frac{\partial}{\partial q^j}+\Delta p_j\frac{\partial}{\partial p_j}\right)\frac{\partial H(\bm{q},\bm{p})}{\partial p^i}\; .
\end{eqnarray}
\end{subequations}
$\lambda$ quantifies the rate of separation of infinitesimally close trajectories, and $\lambda\le0$ in the stable regimes with equality holding for conservative systems. To find  regions of the phase space corresponding to (un)stable dynamics, an exhaustive computation of trajectories, and the corresponding tangent dynamics, is in general necessary.  It is, therefore, desirable to have a method which could predict the stability, from static ``geometrical'' properties.
\subsection{Geometrization of dynamics} 
 Geometrization of dynamics  consists of representing the physical trajectories in time by geodesics on a manifold with a suitable metric. Geometrization has the theoretical upshot in that it provides an alternative framework for analyzing the dynamics. Geometrization does not necessarily provide a computationally efficient advantage over the standard methods, yet it has proved useful in quantifying stability from the curvature of the corresponding metric (see for example \cite{Mont2004}).
 
 Inspired by the general theory of relativity, Eisenhart \cite{Eisenhart} proposed a geometrical description of classical dynamics, for systems which can be described by the least action principle. Eisenhart's metric is defined on the enlarged  configuration space-time with $N+2$ coordinates $x^{\nu}=\lbrace q^0\equiv t, q^i,q^{N+1}: i=1\ldots N\rbrace$, by the differential arc length ($ds$)   
\begin{equation}
ds^2=-2V(\bm{q})(dq^0)^2+a_{ij}dq^{i}dq^{j}+2dq^{0}dq^{N+1}\; .
\end{equation}
\noindent For any geodesic, the extra coordinate  has the solution
\begin{equation}
\qquad q^{N+1}=\frac{\kappa^2}{2}t+C_0-\int_0^t {\cal L}dt'\; ,
\end{equation}
\noindent  where ${\cal L}$ is the Lagrangian, and $C_0$ and $\kappa$ are  arbitrary real constants. With this metric, physical motions satisfy an {\it affine parametrization} $ds^2=\kappa^2 (dt)^2$ (i.e., the  arc length of any geodesic is proportional to the time elapsed).
\subsection{Geometrical version of the stability analysis}
It has been proposed that stability can be quantified based on a lyapunov exponent in a geometrical framework \cite{Pettini1}. Analogous to ${\bm \xi}^{}_T\equiv\Delta q$, one defines the vector field of geodesic spread   
 \begin{equation}
\xi_G^i(s)\equiv \Delta q^i(s) \label{xi}
\end{equation} 
\noindent   as a variation of  geodesics at constant arc length. ${\bm \xi}^{}_G$  satisfies the linearized Jacobi-Levi-Civita (JLC) equations  
\begin{equation}
\frac{D^2\xi^i_G}{ds^2}+R^i_{jkm}\frac{dq^j}{ds}\xi^k_G\frac{dq^m}{ds}=0\; ,\label{JLC}
\end{equation}
\noindent where $\frac{D}{ds}$, and $R^i_{jkm}$ are the covariant derivative along geodesics, and the Riemann-Christoffel curvature tensor \cite{Levi}\footnote{In terms of the Christoffel symbols  $\Gamma^k_{lj}=\frac{1}{2}g^{km}\left(g_{lm,j}+g_{mj,l}-g_{lj,m}\right)$ respectively.  The covariant differentiation and the curvature tensor are defined as
\begin{eqnarray}
D\xi^{i}&=&d\xi^{i}+\Gamma^i_{lj}\xi^{l}dq^j\; ,\nonumber\\
R^i_{jlk}&=&\frac{\partial{\Gamma^i_{jk}}}{\partial{q^l}}-\frac{\partial{\Gamma^i_{jl}}}{\partial{q^k}}+\Gamma^m_{jk}\Gamma^i_{ml}-\Gamma^m_{jl}\Gamma^i_{mk}\; .\label{curve}\nonumber
\end{eqnarray}
\noindent Eq.(\ref{JLC}) can be rewritten as  
\begin{equation}
\frac{d^2\xi^k}{ds^2}+2\Gamma^k_{lj}\frac{dq^l}{ds}\frac{d\xi^j}{ds}+\Gamma^k_{lm,j}\frac{dq^l}{ds}\frac{dq^m}{ds}\xi^j=0\label{JLCopen}\; .\nonumber
\end{equation}}, respectively. The stability is quantified by the geometrical lyapunov exponent \cite{Pettini1}
\begin{equation}
\lambda_G\equiv\lim_{s\rightarrow\infty}\frac{1}{s}\ln\left(\frac{||\bm{\xi}_G(s)||}{||\bm{\xi}_G(0)||}\right)\; .
\end{equation}

This definition formally excludes the variations with respect to the momenta of the initial conditions. Nevertheless, in practice most of the dynamical features pertaining to stability can be explored by variations in coordinates alone. 

With  Eisenhart's metric, the spacial components of Eq. (\ref{JLC}) are equivalent to Eqs. (\ref{TD}) \cite{Reinel}; therefore, $\lambda_G\equiv\lambda$. In words, the Lyapunov exponent obtained with Eisenhart's metric is equal to the one obtained with the tangent dynamics. Furthermore, Pettini et.al. \cite{Pettini4,Pettini5} have shown that the (in)stability of some Hamiltonian systems can be quantified by means of the Ricci curvature, rendering unnecessary the tedious integration of the equations of motion. The formalism based on Eisenhart's ($N+2$)-dimensional space also has the advantage of  being applicable to systems with time-dependent Hamiltonians. 

Several works in the last two decades have been dedicated to show that, in the $N$-dimensional configuration-space, the geometrical-stability analysis can be also carried out employing the kinetic energy metric, also known as the {\it Jacobi} metric \cite{Pettini2,Pettini3,Pettini7} 
\begin{equation}
(g_J)_{ij}\equiv 2[E-V(\bm{q})]a_{ij}(\bm{q})\; .\label{Jacg}
\end{equation} 
The interest in Jacobi's metric was partly due to the fact that the resulting scalar curvature, Ricci's curvature, and sectional curvatures are positive, in many systems of interest \cite{Pettini2,Pettini3,Reinel}. This seemed to support the hypothesis that negative curvature is not the fundamental source of instability, and  that the non-negative oscillating curvature leads the systems to chaos through parametric resonance \cite{Pettini1}.

 In order to test the validity of the geometrical approach within Jacobi's metric, $\lambda^{J}_G$ has been computed for some model Hamiltonians and these results have been compared with those from tangent dynamics. The evidence has not been conclusive. JCL equations within the Jacobi metric are generally cumbersome; in fact, they are tangent dynamics equations with some extra terms having no clear physical interpretation \cite{Pettini7}. Very few examples of exact numerical integration of Eq.(\ref{JLC}) have been presented \cite{Pettini3,Reinel}, and there is no intuitive understanding of the results \cite{Reinel,Pettini7}. An approximate version of Eq.(\ref{JLC}) was used for stability analysis of a large system of self-gravitating particles, and a surprising suppression of chaos was observed,  with increasing number of particles \cite{Pettini2}. This unexpected behavior was assumed to originate from the mathematical approximations that were made for obtaining an equation for the  dynamics of the scalar $||{\bm \xi}^{}_G||$ \cite{Pettini2,Pettini3}. 
 
Szydlowski et.al. \cite{szyd1,szyd2} have already pointed out some inadequacies of the Jacobi metric. A central argument has been that the curvature tensor becomes {\it singular} at the boundaries of the configuration space. There, the kinetic energy vanishes, and there is an infinite number of geodesics, none of which corresponds to a physical trajectory. One may suspect that for many degrees of freedom this framework could give meaningful results, since the probability of reaching the boundaries (i.e., the system coming to a full halt) is very low. However,  this has not been made rigorously. Another limitation of Jacobi's metric is that, since it depends critically on the total energy ($E$), only those variations that do not change $E$ must be considered.

In this paper we show that the geodesic spread and the tangent dynamics fields are equivalent when the geodesic length is only a function of time (if any  pair of geodesics, which cover equal  time intervals, have equal arc length).   We apply the formalism to  a system at the onset of chaos, and to a trivially stable system (i.e., system of harmonic oscillators). In these cases we found that Jacobi's metric predicts an unstable dynamics for stable systems,   even for $N\gg 2$, where the kinetic energy does not vanish. The  geodesic-spread within the Jacobi metric  suffers  from non-physical  parametric resonance, and that this is related to the fluctuations of the {\it kinetic energy}. 

As such, endowing the configuration space with Jacobi metric does not always provide an appropriate measure for the calculation of the Lyapunov exponent, and the consequent analysis of dynamical instability.  The manuscript is organized as follows. In section \ref{thetwoxi} the relationship between tangent dynamic and geodesic spread, and the differences in the corresponding equations are derived and discussed.  In section \ref{Morsepot} stability analysis by means of Jacobi metric is compared to the results from tangent dynamics for a two-dimensional system. In section \ref{harmonic} a similar comparison is made employing a stable and analytically soluble system (harmonic oscillators). General conclusions are drawn in section \ref{conclus}.

\section{Relationship between ${\bm \xi}^{}_T$ and ${\bm \xi}^{}_G$}\label{thetwoxi}

A variation ($\Delta q^i_{}$) of the coordinates  $q^i$ (measured between two trajectories)  can be written as  
\begin{equation}
\Delta q^i_{}=\xi^i_T+\dot{q^i}\Delta t \label{relation1},
\end{equation}

\noindent where $\xi^i_T$ is the variation at a fixed time, and the second term accounts for a  time variation.  Let the arc length of a geodesic be parametrized by
\begin{equation}
s=\int_0^T  F({\bm q},\dot{{\bm q}}) dt.
\end{equation}

The  difference in the arc length of two trajectories, up to first order in coordinates and time variations, is 
\begin{widetext}
\begin{equation}
\Delta s=\left(F-\dot{q}^i\frac{\partial F}{\partial \dot{q}^i}\right)\Delta t\vert^{}_T+\frac{\partial F}{\partial \dot{q}^i}\Delta q^i_{}\vert^{}_T-\frac{\partial F}{\partial \dot{q}^i}\Delta q^i_{}\vert^{}_0+\int_0^T \left(\frac{\partial F}{\partial q^i} -\frac{d\phantom{a}}{dt}\frac{\partial F}{\partial \dot{q}^i}\right) \xi^i_T dt\; . \label{deltaSS}
\end{equation}
\end{widetext}
The r.h.s. integral vanishes along any geodesic because the terms in parenthesis are the equations of motion. Remember that the latter are obtained from the extremal condition ($\Delta s=0$) with respect to variations that keep the boundary conditions unchanged. Thus, the $\Delta s$ between two physical trajectories with slightly different initial conditions is 
\begin{equation}
\Delta s=\left(F-\dot{q}^i\frac{\partial F}{\partial \dot{q}^i}\right)\Delta t\vert^{}_T+\frac{\partial F}{\partial \dot{q}^i}\Delta q^i_{}\vert^{}_T-\frac{\partial F}{\partial \dot{q}^i}\Delta q^i_{}\vert^{}_0\; ; \label{deltaSSS}
\end{equation}
\noindent and it simplifies to different forms depending on the variational formalism. In geometrical formalisms with trajectory-independent arc length (such as Eisenhart's approach), the difference in the arc length of two neighboring geodesics reduces to
\begin{equation}
\Delta s=F\Delta t\; ,
\end{equation}
\noindent where $F$ turns out to be constant.  As such,  variations at constant $t$ correspond to variations at constant $s$. In variational formalisms where $s$ is minimal with respect to variations with unconstrained time (e.g., within Jacobi metric), the term multiplying $\Delta t\vert_T^{}$  vanishes; i.e.,
\begin{equation}
F-\dot{q}^i\frac{\partial F}{\partial \dot{q}^i}\equiv 0 \label{conserv}.
\end{equation}
\noindent Remember that in the case of the Jacobi metric $F={\cal L}+E$, and variations are taken at constant energy $E$, with unconstrained time.  In such cases, $s$ is trajectory dependent, and the difference in the arc-length of  neighboring geodesics is 
\begin{equation}
\Delta s=\frac{\partial F}{\partial \dot{q}^i}\Delta q^i_{}\vert^{}_T-\frac{\partial F}{\partial \dot{q}^i}\Delta q^i_{}\vert^{}_0\label{NAP}\; .
\end{equation} 
\noindent Consequently, a geodesic spread at constant  $s$,  $\xi^i_G\equiv\Delta q^i_{}\vert^{}_{\Delta s=0}$, is only possible if  $\frac{\partial F}{\partial \dot{q}^i}\xi^i_G$ is constant in time, or equivalently if  
\begin{equation}
\frac{d\phantom{a}}{dt}\left(\frac{\partial F}{\partial \dot{q}^i}\xi^i_G\right)=0\label{dsnull}\; .
\end{equation}
\noindent If we assume that Eq.(\ref{dsnull}) can be fulfilled, then, one can multiply Eq.(\ref{relation1}) by $\frac{\partial F}{\partial \dot{q}^i}$ and find the corresponding time mismatch $\Delta t$ between the two geodesics with equal arc, and we arrive at the following transformation  between ${\bm \xi}^{}_G$ and ${\bm \xi}^{}_T$:
\begin{equation}
{\bm \xi}^{}_G={\bm M}{\bm \xi}^{}_T+C\left(\frac{\partial F}{\partial \dot{q}^i}\dot{q}^i\right)^{-1}\dot{\bm q}\; ,
\end{equation}
\noindent where  $C=\frac{\partial F}{\partial \dot{q}^i}\xi^i_G$, and 
\begin{equation}
M^j_i=\delta^j_i-\frac{\partial F}{\partial \dot{q}^i}\dot{q}^j/\frac{\partial F}{\partial \dot{q}^k}\dot{q}^k .
\end{equation}
\noindent It is easy to show that ${\bm M}^2={\bm M}$, and Tr(${\bm M})=N-1$; thus,  ${\bm M}$ is a projector of rank $(N-1)$. The tangential component of ${\bm \xi}^{}_G$ is not well-defined in the above relations. At first sight, one could think that a suitable choice of coordinates might isolate this component. However, the geodesic spread at constant arc length turns out to be ill-defined. Eq.(\ref{dsnull}) cannot be fulfilled and the constant $C$ does not exist. To show this, we proceed as follows.

 Carrying out the differentiation in Eq.(\ref{dsnull}), it takes the form 

\begin{equation}
 \xi^i_G\frac{d\phantom{a}}{dt}\left(\frac{\partial F}{\partial \dot{q}^i}\right)+\frac{\partial F}{\partial \dot{q}^i}\dot{\xi}^i_G=0\; ;
\end{equation} 
\noindent which, together with the equations of motion, give 

\begin{equation}
\xi^i_G \frac{\partial F}{\partial q^i}+ \frac{\partial F}{\partial \dot{q}^i}\dot{\xi}^i_G =0 \label{dFconsT}\; .
\end{equation} 
 At the same time, since Eq.(\ref{conserv}) is an identity that has to be fulfilled by any geodesic,  its linearized variation must be identically zero; i.e.,
 \begin{eqnarray} 
 0&\equiv & \Delta\left(F-\dot{q}^i\frac{\partial F}{\partial \dot{q}^i}\right)\nonumber\\
 &=&\frac{\partial F}{\partial q^i}\xi^i_G-\dot{q}^i\Delta \left(\frac{\partial F}{\partial \dot{q}^i}\right)\; \label{contr}.
 \end{eqnarray}
 \noindent Satisfying both Eq.(\ref{contr}) and Eq.(\ref{dFconsT}) would require that 
 
 \begin{equation}
 \Delta \left(\dot{q}^i\frac{\partial F}{\partial \dot{q}^i} \right)=0\; .\label{Strong}
 \end{equation}
 Eq.(\ref{Strong}) is a very strong condition which, within the Jacobi metric, means that the variations at constant arc length must also leave the kinetic energy unchanged.  For the sake of simplicity, and since it does not affect the generality of the analysis, let us repeat the above derivation for the Jacobi metric  assuming that the elements of the inertia matrix ($a_{ij}^{}$) are constant.
 
  The fact that the Jacobi metric depends critically on $E$ requires that we restrict the variations to a constant energy hypersurface. This is actually  equivalent to Eq.(\ref{contr}); and from this condition, the variations must satisfy
\begin{equation}
a_{ij}^{}\dot{q}^i\dot{\xi}^j_{G}+V^{}_{,k}\xi^k_{G}=0\; .\label{constr1cart}
\end{equation}   

\noindent  On the other hand, Eq.(\ref{dFconsT}) gives
\begin{equation}
a_{ij}^{}\dot{q}^i\dot{\xi}^j_{G}-V^{}_{,k}\xi^k_{G}=0\; .\label{constr2cart}
\end{equation}

\noindent The two constraints (Eq.(\ref{constr1cart}) and Eq.(\ref{constr2cart})) require that ${\bm \xi}$ leaves both total energy and kinetic energy unchanged at any time. In other words, $\dot{\bm \xi}$ has to be orthogonal to the momentum, and ${\bm \xi}$ orthogonal to the force.  We will show now that these two conditions are incompatible. But first, let us show that the tangent dynamics does allow ${\bm \xi}$ to remain in the constant-energy hypersurface.

 The equations of the tangent dynamics, for constant $a_{ij}^{}$, are the set of equations
\begin{equation}
\ddot{\xi}_T^n+a^{nk}V_{,kl}\xi^l_T=0 \; .\label{TD1}
\end{equation}
\noindent  The set Eq.(\ref{constr1cart}) constrains the variation to a constant energy hypersurface. The dynamics of ${\bm \xi}$ must allow this to hold over time; therefore the time derivative of Eq. (\ref{constr1cart})  must be identically zero.  By performing this derivative, and taking Hamilton's equations into account, one obtains 
\begin{equation}
p^{}_n\left(\ddot{\xi}_{T}^n+a^{nk}V_{,kl}\xi^l_{T}\right)=0 \; .\label{dE}
\end{equation}
\noindent Because of Eq.(\ref{TD1}), all coefficients multiplying the momenta in Eq. (\ref{dE}) are zero, and therefore the above condition is satisfied for any physical trajectory. Within the Jacobi metric, this is not the case. The corresponding JLC equations  are  
\vspace{-10 mm} 
\begin{widetext}
\begin{equation}
\ddot{\xi}_G^n+a^{nk}V_{,kl}\xi_G^l=-\frac{1}{T}\left( a^{nm}V_{,m}\lbrace a_{ij}\dot{q}^i\dot{\xi}^j_G+V_{,l}\xi^l_G\rbrace-\dot{q}^n\lbrace V_{,il}\dot{q}^i\xi^l_G+V_{,j}\dot{\xi}^j_G+\frac{1}{T}V_{,i}\dot{q}^iV_{,l}\xi^l_G\rbrace\right)\label{Jacdyn}\; ,
\end{equation}
\end{widetext}
\noindent  where $T=E-V$. If we could guarantee that the  variations (${\bm \xi}^{}_G$ and $\dot{\bm \xi}^{}_G$) do not alter the total energy, the first term in curly braces  would vanish, and the equations would reduce to
  \begin{equation}
\ddot{\xi}_G^n+a^{nk}V_{,kl}\xi_G^l=\frac{\dot{q}^n}{T}\left( V_{,il}\dot{q}^i\xi^l_G+V_{,j}\dot{\xi}^j_G+\frac{1}{T}V_{,i}\dot{q}^iV_{,l}\xi^l_G\right)\; .\label{JacdynEconst}
\end{equation}
Comparing  Eqs.(\ref{TD1}) with Eqs.(\ref{JacdynEconst}), we see that the latter contains extra terms (the r.h.s.). These terms can be rewritten as $\dot{q}^n\frac{d\phantom{a}}{dt}((E-V)^{-1}\Delta V)$. Thus,  the differential equations for ${\bm \xi}_G$ would become identical to the equations of the tangent dynamics only when
\begin{equation}
\frac{d\phantom{a}}{dt}\left(\frac{V^{}_{,k}\xi^k_G}{E-V}\right)\equiv 0\; .\label{wrong}
\end{equation}
\noindent  However, Eq.(\ref{wrong}) is not protected by the dynamics derived from Eqs.(\ref{JacdynEconst}). Namely, this equality cannot be derived from the equations of motion. As a result, even if we choose initial conditions for  ${\bm \xi}^{}_G$ and $\dot{\bm \xi}^{}_G$, which satisfy  $\delta T=\delta E=0$,   Eq.(\ref{wrong}) will not hold over time. 

The work of Sospedra and co-workers\cite{Reinel} apparently shows that, in contrast to our results, the component of ${\bm \xi}^{}_G$ in the direction of the trajectory can be decoupled from the system of equations, and it does not accelerate. However, their result is a consequence of replacing the covariant derivative of a quantity that is not a true scalar (the projections of ${\bm \xi}^{}_G$ on a given basis) by a simple derivative. 

We will show numerically in the next sections that the extra terms  in Eq. (\ref{JacdynEconst}) introduce non-physical instabilities. But let us first make a simple heuristic analysis to get the feeling of the problem. As before, we take the tangent dynamics as a reference of validity to compare against.

 When the energy of a system is close enough to its  value at a minimum of the potential, the motion is restricted to small periodic oscillations around this equilibrium point; the dynamics is stable. We can expand the potential up to second order in ${\bm q}$. Then, all $V_{,kl}$ are  constant, and the tangent dynamics (from Eq.(\ref{TD1})) gives oscillatory solutions for $\xi_T$ (therefore, $\lambda=0$), in correspondence to a dynamically  stable system. If we consider now trajectories with higher energy, higher orders  in the ${\bm q}$ dependence of $V$ must be taken into account.  This may drive the system towards chaotic behavior.  $V_{,kl}$  are ${\bm q}$-dependent, and oscillate at harmonics of the frequencies of the system. They may produce parametric resonance in the tangent dynamics, and result in $\lambda>0$ \cite{Landau}. Thus, parametric resonance in $\xi$-dynamics is relevant at the on-set of chaos \cite{Pettini1}.  For even higher energies, the trajectories may cross hyperbolic points between the minima of the potential; the system may visit most regions of the phase space, and the trajectories are no longer periodic. Now we repeat the analysis using Jacobi metric. 
 
Because of the first order derivatives ($V_{,j}$), and the $\dot{\bm q}$-dependence of the  the r.h.s. of Eqs.(\ref{JacdynEconst}), the coefficients in these linear equations are time-dependent, even  when $V({\bm q})$ is quadratic in ${\bm q}$. These terms oscillate at frequencies that are harmonics of the fundamental modes of the system, and so they may create parametric resonance \cite{Landau}, resulting in positive $\lambda^{}_G$'s for a stable system. At the actual onset of chaos, when higher orders in the ${\bm q}$-dependence of $V$ are important,  real parametric resonance appears from the second term on the l.h.s.. It is not possible to discriminate the false exponential divergence from the physical one, when the exponent is computed from the evolution of ${\bm \xi}_G$ according to Eqs.(\ref{JacdynEconst}). A positive exponent may be obtained in both, stable and unstable regimes.

\section{A two-dimensional example. Physical and unphysical instabilities}\label{Morsepot} 

The first representative example that we take is a two-dimensional system described by the Hamiltonian 
\begin{equation}
H=\frac{P^2_R}{2\mu_1}+\left(\frac{1}{2\mu_1R^2}+\frac{1}{2\mu_2r_e^2}\right)P^2_{\theta}+V(R,\theta)\; .
\end{equation}
\noindent  This represents the energy of a particle moving with respect to the center of mass of a rigid dimer of length $r_e=3.0271\AA$, when the total angular momentum is $J=0$; it has been the subject of investigation due to its interest for molecular dynamics \cite{Reinel,Ronce,Gut1,Gut2,Gut3,Gut4,Rub}. $R$ and $\theta$ are polar coordinates, with polar axis running along the dimer. The interaction between the particle and each member of the dimer is represented by a Morse potential $V(r)=D\left[1-{\text e}^{-\alpha (r- d)}\right]^{2}_{}$, with parameters $D=40.75$ cm$^{-1}$, $\alpha=1.56 \AA^{-1}$, and $d=4.36\AA$ \cite{Reinel}.  The resulting  potential surface has two equivalent minima at ($R=\sqrt{r_e^2/4+d^2_{}}$; $\theta=\pm\pi/2$). These minima are connected by two equivalent trenches going around the dimer, with saddle points at ($R=d+\frac{1}{\alpha}\ln(\cosh(\alpha r^{}_e))-\frac{1}{\alpha}\ln({\cosh(\alpha r^{}_e/2)})$; $\theta=0,\pi$). The relative height  $E^{}_{\text{saddle}}-E^{}_{\text{min}}=2D[1+\coth^2_{}(\alpha r^{}_e/2)]^{-1}_{}$ is  $\approx 40.66$ cm${}^{-1}_{}$. The reduced masses are taken as $\mu_1=18$ amu, and $\mu_2=64$ amu \cite{Reinel}. In the following, all relevant quantities  are expressed in powers of cm. Energy, linear momenta, and  $\lambda$ are in cm$^{-1}$, time and distances are in cm and angular momentum is dimensionless.
\begin{figure}[htb]
\centering
\includegraphics[width=0.8\linewidth]{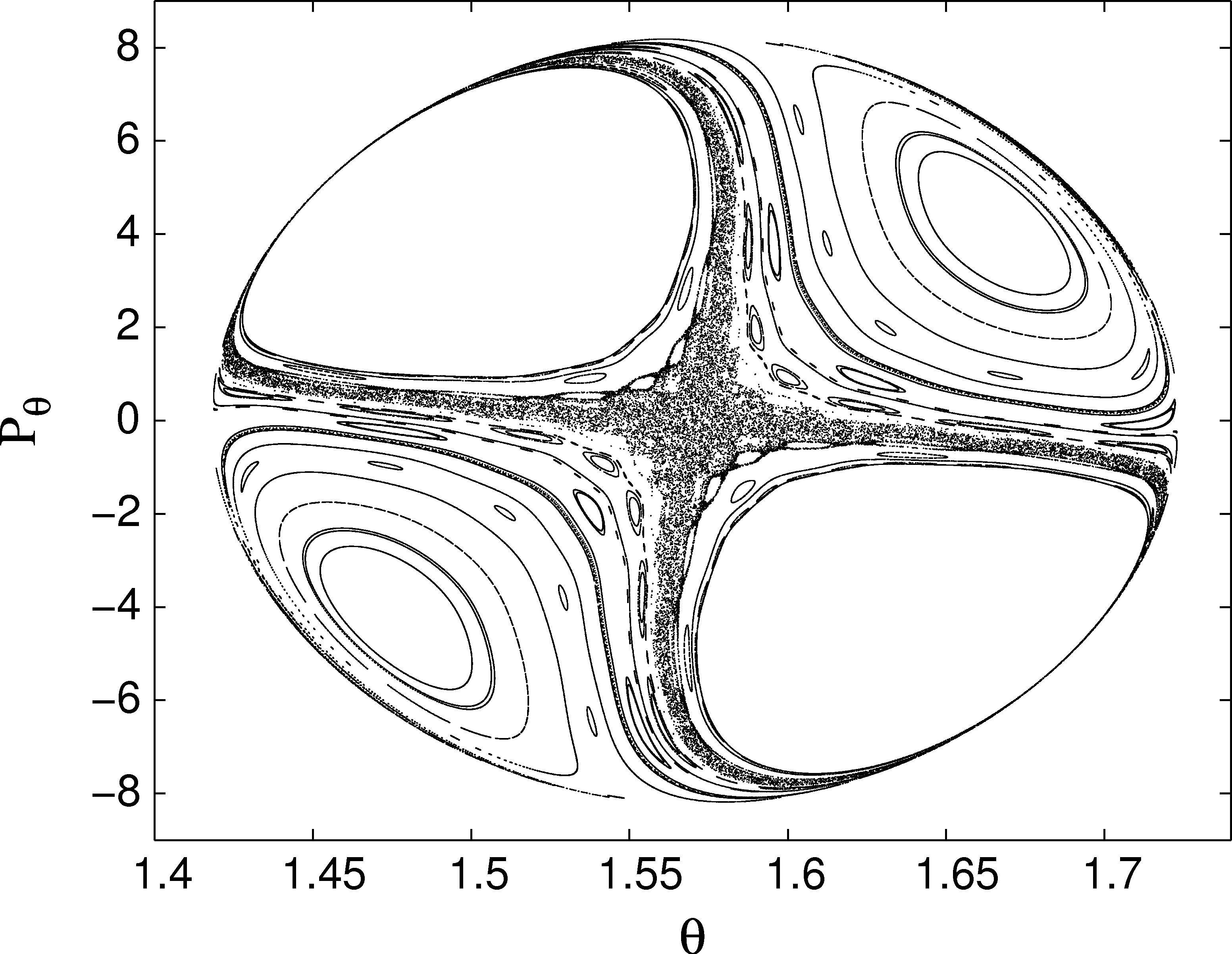}
\caption{\label{fig1} Poincare surface section for $\Delta E = 6.5$}
\end{figure}

 We have performed an exhaustive integration of trajectories, and we have analyzed Poincare surface sections (PSS) for different values of the total energy.   We present in  Fig.\ref{fig1} the PSS $\theta$-$P^{}_{\theta}$, for $\Delta E = E - E^{}_{\text{min}}= 6.5$ cm$^{-1}$. At this energy, there are wide regions that contain stable limit-cycles, and there are also regions of unstable dynamics. We chose initial conditions at the edge of a stable region, and computed the exponents ($\lambda_G^{}$, and $\lambda$), for $\Delta E= 6.5$ cm$^{-1}$, and for several energies above and below. This should allow us to observe how the on-set of chaos is detected by the geometrical exponent, and to compare it with the results from tangent dynamics. 
\begin{figure}[htb]
\centering
\includegraphics[width=0.8\linewidth]{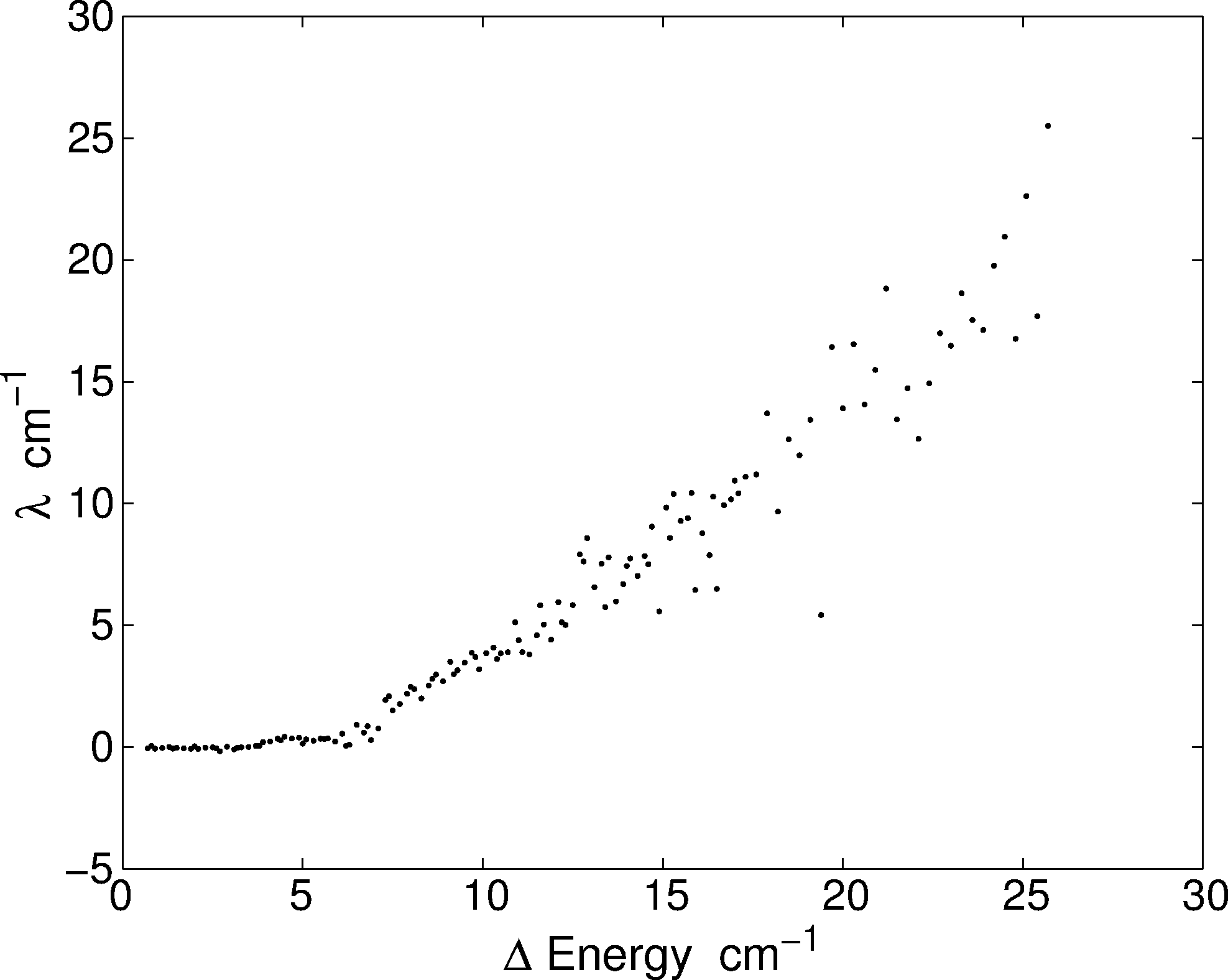}
\caption{\label{fig2} The Lyapunov exponent obtained from the tangent dynamics of trajectories at the edge between stable and unstable regions in Fig.\ref{fig1}, for several values of the total energy.}
\end{figure} 
\begin{figure}[htb]
\centering
\includegraphics[width=0.8\linewidth]{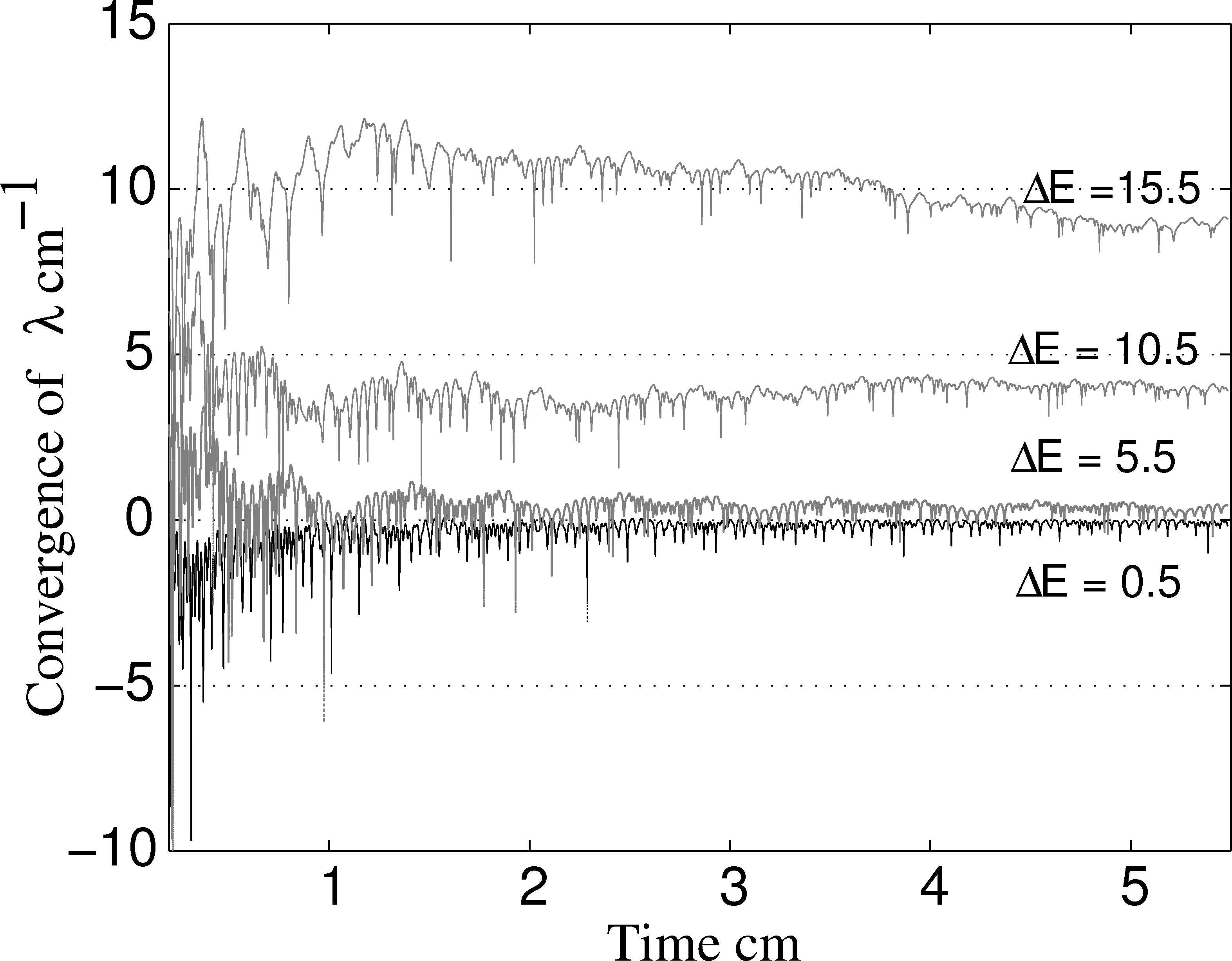}
\caption{\label{fig3} The convergence of the Lyapunov exponent for different energies using Eisenhart metric (tangent dynamics).}
\end{figure}
\begin{figure}[htb]
\centering
\includegraphics[width=0.8\linewidth]{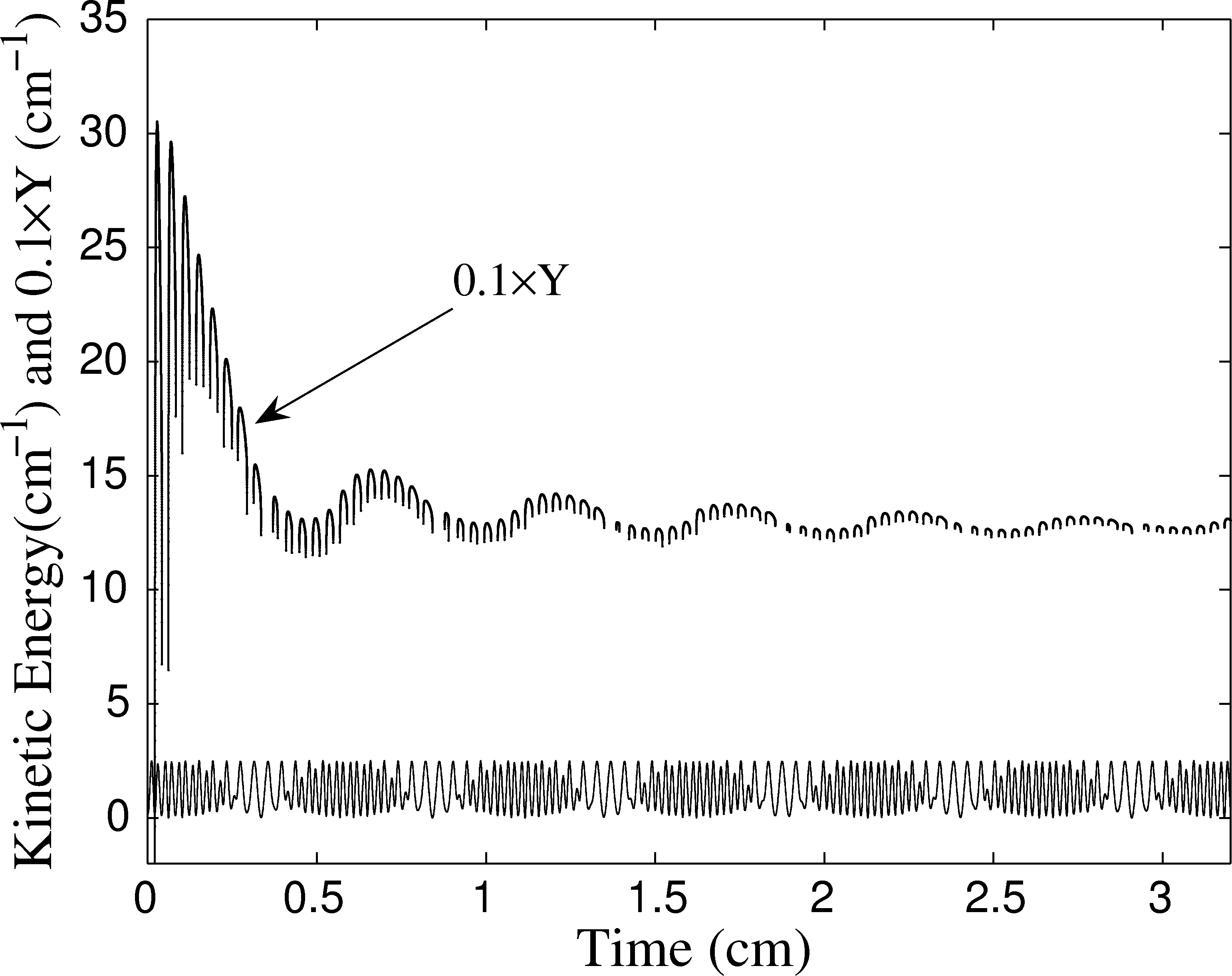}
\caption{\label{fig4} Kinetic Energy and ``convergence'' of the Lyapunov exponent for $\Delta E$=2.5 cm$^{-1}$ using Jacobi metric.}
\end{figure}
\begin{figure}[h!]
\centering
\includegraphics[width=0.8\linewidth]{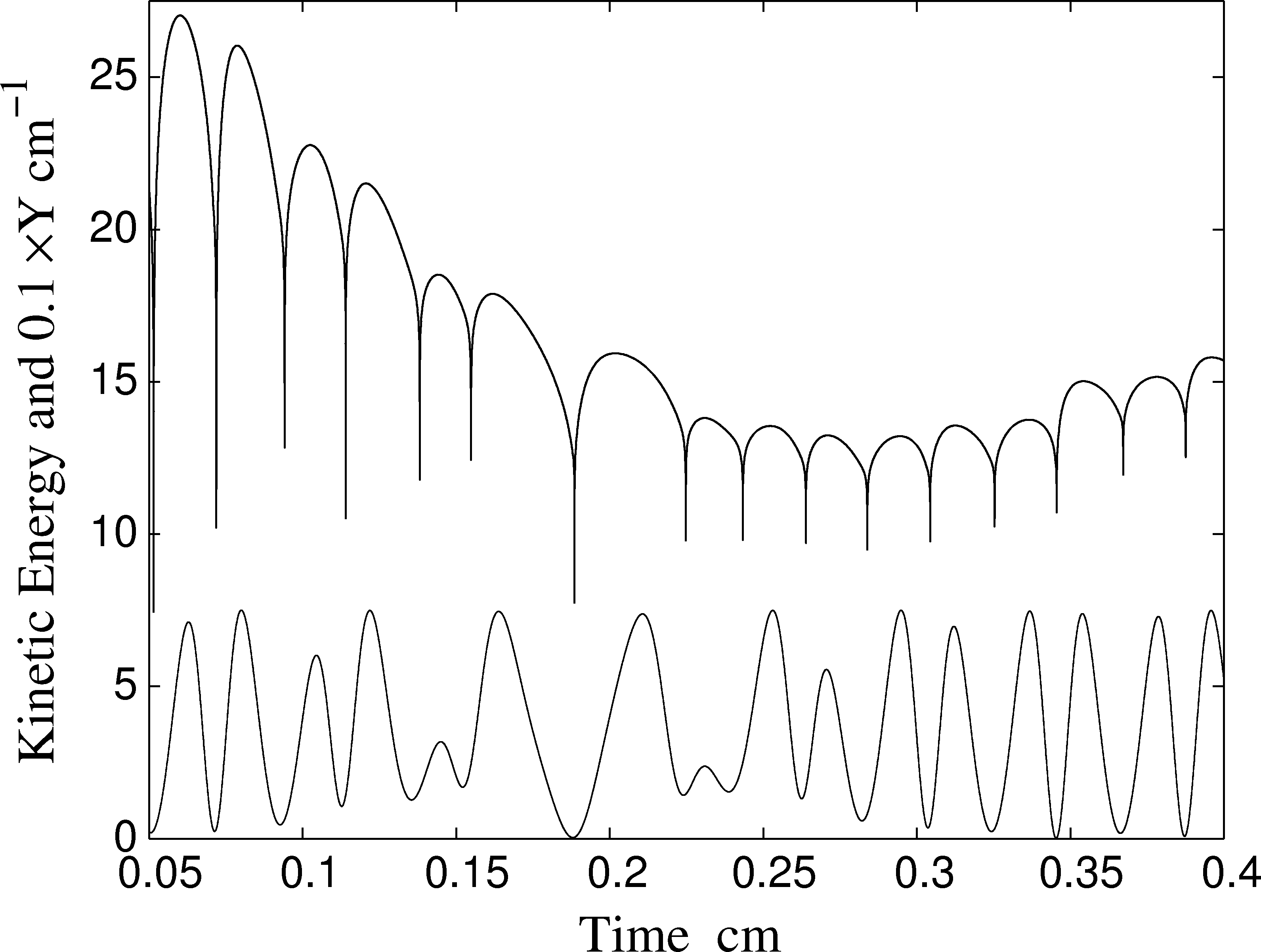}
\caption{\label{fig5} Zooming in on Fig.\ref{fig4}: Kinetic Energy and the corresponding ``kicks'' in $\xi$ for $\Delta E$=2.5 cm$^{-1}$.}
\end{figure}

We plot the Lyapunov exponent obtained with the Eisenhart metric (actually,  tangent dynamics) versus the total energy, in Fig.\ref{fig2}.   The on-set of chaos seems to take place around  $\Delta E = 4.3$ cm$^{-1}$, as indicated by positive values of $\lambda$. For a reason that will become clear soon, it is interesting to look at the early-time evolution of   $Y(t)=\frac{1}{t}\log(\frac{||\xi(t)||}{||\xi(0)||})$  for different energies, ranging from very stable ($\Delta E=0.5$ cm$^{-1}$) to unstable   ($\Delta E=15.5$ cm$^{-1}$) regimes. 
Fig.\ref{fig3} shows the stabilization of $Y$ around the corresponding limiting value. We intended to calculate the Lyapunov exponent  using Jacobi metric, and we faced serious memory overflows, due to a explosive growth of $\|{\bm \xi}_G^{}\|$. Its time evolution  is exponential even for trajectories in the stable regime (e.g., $\Delta E=2.5$ cm$^{-1}$), as it can be seen in Fig.\ref{fig4}. We extracted from the $Y$ dynamics a very short time window which is shown in  Fig.\ref{fig5}. In contrast to the results from tangent dynamics, one can see the "kicks" produced in the evolution of $Y$, associated to the oscillations of the kinetic energy. $Y$ converges to about 130 cm$^{-1}$; in contradiction to the vanishing exponent obtained with Eisenhart metric, and also in contradiction with PSS calculated in the corresponding energy range.

These results gave us the idea to look at a model where one could quantify the oscillations of the kinetic energy with a minimal error and at minimal cost, in  order to search for a correlation between the positive $\lambda^{}_G$ and the fluctuations of $T$.  This is the subject of the next section.

We like to comment here that during the review of this manuscript we became aware\footnote{We thank Prof. R. Montgomery for having kindly suggested these valuable references within the review of our manuscript.} of other results obtained by Motter and co-workers, which are in good agreement with ours \cite{Motter1,Motter2,Motter3}. They found that the Lyapunov exponent could be a reliable indicator of stability as long as the time-reparametrization does not create singularities in the invariant measure. In turn, singular reparametrizations  can shift the Lyapunov exponent. The Jacobi metric is actually an example of singular reparametrizations. In a very recent work \cite{Mont} it has been also shown that, within the Jacobi metric, there are trajectories which fail to minimize the arc length.

\section{ ${\bm \lambda^{}_G}$ from Jacobi metric    {\em vs}  fluctuations of the kinetic energy.}\label{harmonic}

Here we use a paradigm of stability:  a system of independent harmonic oscillators. We apply the stability analysis using Jacobi metric; any evidence of chaos can then be understood as a failure of the method. We investigate  whether there is a relation between $\lambda^{}_G$ and the fluctuation of the kinetic energy. 

The time-independent Hamiltonian 
\begin{equation}
H(\bm{q},\bm{p})=\frac{1}{2}\left(\delta^{ij}p_{i}p_{j}+ \omega^2\delta_{ij}q^iq^j\right)\; , \label{Hamilton}
\end{equation}
\noindent with $p_{i}=\delta_{ik}\dot{q}^k$ is our basic model. The solutions to the equations of motion are
\begin{equation}
q^k(t)=C^k\cos(\omega t+\theta_k)\; , \label{dysol}
\end{equation}
\noindent where $C^k$ and $\theta_k$ depend on the initial conditions, and we chose them  as $C^k=1$, and $\theta_k=k\frac{2\pi f}{N}$, $k=1,...,N$. As before, $N$ stands for the number of degrees of freedom. The phases, $\theta_k$, are homogeneously distributed on a fraction ($f$) of $2\pi$ (phase circle). These settings allow us to find an analytical expression for the fluctuation $\sqrt{\sigma}$, and to tune it varying the values of $f$ and $N$. Using Eq.(\ref{dysol}), we obtain for $T$, and for its normalized variance ($\sigma$) 
\begin{eqnarray}
T&=&N\left(\frac{\omega C}{2}\right)^2\left[1-\sqrt{2\sigma}\cos(2\omega t+2\pi f\frac{N+1}{N})\right]\; , \\
\sigma&\equiv&\frac{\langle T^2\rangle-\langle T\rangle ^2}{\langle T\rangle^2}=\left(\frac{\sin(2\pi f)}{\sqrt{2}N\sin(2\pi f/N)}\right)^2\; .\label{fluct}
\end{eqnarray}

 $\sigma$ decreases  with increasing $N$, having  the limit $\sigma\rightarrow |\frac{\sin(2\pi f)}{2\pi f\sqrt{2}}|$ as $N\rightarrow\infty$. When all of the oscillators are in phase ($f=0$), $\sigma$ takes its maximum value and $T$ becomes  zero  every $\Delta t=\frac{\pi}{\omega}$. 

 JLC equations with Eisenhart metric (and the equations of the tangent dynamics) take the form
\begin{equation}
\frac{d^2\xi^k_T}{dt^2}+\omega^2\xi^k_T=0\; ,\label{eisen}
\end{equation}
\noindent whereas with the Jacobi metric,
\vspace{-10 mm} 
\begin{widetext} 
\begin{equation}
\frac{d^2\xi^k_G}{dt^2}+\omega^2\xi^k_G + \frac{\omega^2}{T}\delta_{lj}\left[\left(q^k{\dot q}^l-q^l{\dot q}^k\right)\frac{d\xi^j_G}{dt}+\left(\omega^2q^kq^l-{\dot q}^k{\dot q}^l-\frac{\omega^2}{T}\delta_{im}{\dot q}^iq^m{\dot q}^kq^l\right)\xi^j_G\right]=0\; . \label{xidyn}
\end{equation}
\end{widetext}
\noindent  Eq.(\ref{eisen})  gives $\lambda=0 $ for any  initial condition. On the other hand, when trajectories (\ref{dysol})  are substituted in Eq.(\ref{xidyn}), the latter take the form
\begin{equation}
\frac{d^2\xi^k_G}{dt^2}+\omega^2\xi^k_G + \omega I^k_j\frac{d\xi^j_G}{dt}+\omega^2\left[J^k_j+K^k_j\right]\xi^j_G=0\; , \label{fin}
\end{equation}
\noindent with the couplings $I^k_j$, $J^k_j$ and $K^k_j$ given by    
\begin{subequations}
\begin{eqnarray}
 I^k_j&=&\frac{\omega^2 C^2}{T}\sin(\theta_k-\theta_j)\; ,\\
 J^k_j&=&\frac{\omega^2 C^2}{T} \cos(2\omega t+\theta_k+\theta_j)\; ,\\
 K^k_j&=&-\frac{\omega^4 C^4}{2T^2}\sin(2\omega t +2\pi f\frac{N+1}{N})\times\\
&&\left[\sin(2\omega t+\theta_k+\theta_j)-\sin(\theta_k-\theta_j)\right]\; . \nonumber
\end{eqnarray} 
\end{subequations}
\noindent Eqs.(\ref{fin}) have a typical  structure from which  parametric resonance may arise \cite{Landau}. The basic frequency of the oscillators is one half of the frequency at which the terms of the ``restoring force matrix'' oscillate (which is the ideal condition for a first order parametric resonance). However, since the components  $\xi^k_G$ have become coupled through the terms $I^k_j$, $J^k_j$ and $K^k_j$, one could still doubt whether a normal mode resonates with the fluctuating parameters.  

\begin{figure}[htb]
\centering
\includegraphics[width=0.8\linewidth]{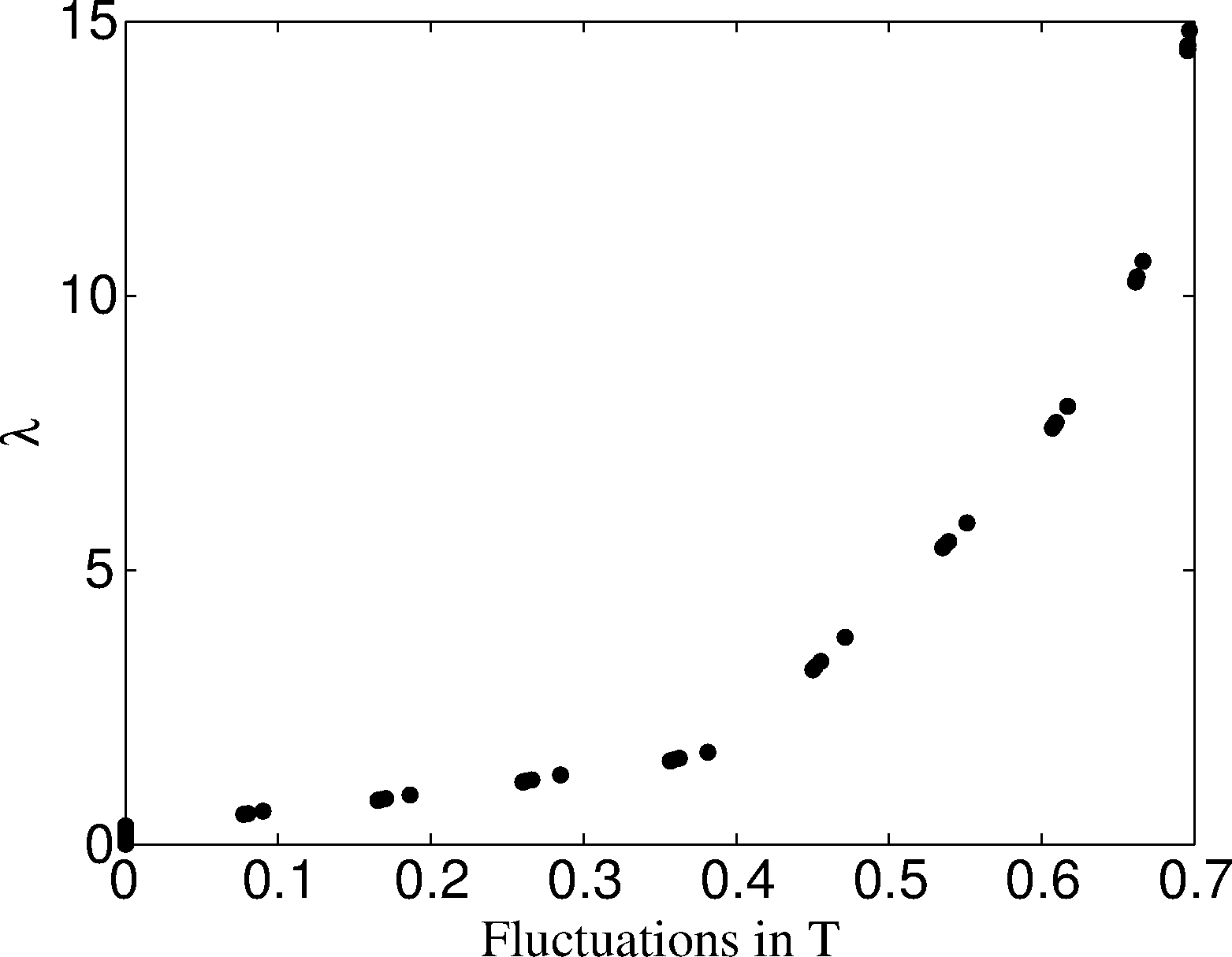}
\caption{\label{fig7}  $\lambda^{}_G$ vs. $\sqrt{\sigma}$ for $ N=2+j^2$ with  $j=1,\ldots,14$ and $f=0.05,0.1,\ldots,0.45$.} 
\end{figure}

Despite the simplicity of the chosen system, the analytical integration of these equations seems quite challenging. Therefore, we leave the analytical evaluation of the Lyapunov exponent for future work. We obtained $\lambda^{}_{G}$  from the numerical integration of Eq.(\ref{fin}),  using several values of $f$, and $N$. In Fig.\ref{fig7} we show $\lambda^{}_G$ {\em vs} $\sqrt{\sigma}$. Note that, although we varied $N$ and $f$ independently, and $\sigma$ depends on both of these parameters, we obtained a smooth curve for $\lambda^{}_G$ vs. $\sqrt{\sigma}$.  This confirms our hypothesis that, within Jacobi metric, $\lambda^{}_G$ grows with $\sigma$, and it does not measure the actual stability of the physical system.  Actually, un-physical parametric resonance is not the only manifestation of the failure of this methodology. The scalar curvature (${\cal K}$) of the manifold is

\begin{equation}
{\cal K}=\frac{N-1}{8(E-V)^3_{}}\left[4(E-V)\nabla^2_{} V -(N-6)| \nabla V|^2_{}\right]\; ,\label{Curv}
\end{equation} 
\noindent which takes negative values in some regions of the accessible configuration space, for any system with $N>6$. For our simple example of harmonic oscillators, the curvature can be rewritten as
\begin{equation}
{\cal K}=\frac{\omega^2_{}(N-1)}{4(E-V)^3_{}}\left[2N(E-V)-(N-6)V\right]\; ,
\end{equation} 
\noindent  which, for $N$ much greater than 6, will be negative at any point with $V> 2E/3$. Eq.(\ref{Curv}) makes evident that the sign of ${\cal K}$ is not an indicator of the  (in)stability of the underlying dynamical system.  We would like to finish this discussion showing a special (and illustrative) case of the above example, where the differential equations for ${\bm \xi}^{}_G$ are simple; the solutions are quite telling  by mere inspection. 

Let us take two identical one dimensional simple harmonic oscillators with coordinates $x$ and $y$, respectively. Take the initial conditions such that they both have non-vanishing amplitudes, and a phase difference of $\frac{\pi}{2}$. With this choice, the kinetic energy never vanishes. This is formally equivalent to a single two dimensional oscillator with a central potential $V(r)=\frac{1}{2}\omega^2_{}r^2_{}$. The orbits are ellipses; a circle corresponding to the special case where both variables oscillate with equal amplitude. The circular trajectory has a constant kinetic energy.   This example also allows a transparent way to control the initial conditions of ${\bm \xi}^{}_G$, so to keep the total energy unchanged (which has been said to be a key to obtain a meaningful stability analysis). The  trajectories can be parametrized in polar coordinates as
\begin{eqnarray}
r^2_{}(t)&=&R^2_{}+\Delta^2_{}\cos(2\omega t)\; , \\
{\dot{\theta}}^2_{}(t)&=&r^{-4}\omega^2_{}(R^4_{}-\Delta^4_{})\; ,
\end{eqnarray}
\noindent with the total energy $E$ and the angular momentum $L$ given by
 \begin{eqnarray}
E&=&R^2_{}\omega^2_{}\;,\\
L^2_{}&=&\omega^2_{}(R^4_{}-\Delta^4_{})\;.
\end{eqnarray}
\noindent $R^2_{}$, and $\Delta^2_{}$ are  the average of $r^2_{}(t)$, and its oscillation amplitude, respectively. The latter is directly related to the fluctuations of the kinetic energy, since   $T=\frac{\omega^2_{}}{2}[R^2_{}-\Delta^2_{}\cos(2\omega t)]$.  The constants satisfy $\Delta^2_{}\leq R^2_{}$, with $\Delta=0$ corresponding to the circular orbit, and $\Delta=R$ to the undesired one dimensional case.  By making use of conservation laws, one can reduce the two coupled equations of ${\bm \xi}^{}_G$ to a single variable problem.

Eisenhart's metric, which is equivalent the tangent dynamics, gives 
\begin{equation}
\ddot{\xi}_T+ \left(V_{,rr}+\frac{3L^2_{}}{r^4_{}}\right)\xi_T =0 \label{Ecent}\; .
\end{equation}
\noindent One can show, by differentiation of Newton's equation, that the solutions of eq.(\ref{Ecent}) are proportional to $\dot{r}(t)$, and are stable,  as expected. For the harmonic potential, one can also show that $\xi_T\propto \left(r-(2L^2_{}/E)r^{-1}_{}\right)$.

On the other hand, in polar coordinates, the components of the metric tensor for the Jacobi metric are $g^{}_{rr}=2[E-V]$, $g^{}_{r\theta}=0$, and $g^{}_{\theta\theta}=2[E-V]r^2_{}$. The differential equation for the radial component of $\xi$ is
\begin{eqnarray}
\ddot{\xi}_G+ \left(V_{,rr}+\frac{3L^2_{}}{r^4_{}}\right)\xi_G &=& \left\{V_{,rr}\left(2-\frac{L^2_{}}{r^2_{}(E-V)}\right)\right . \label{Jcent}\\
&+ & V_{,r}^2\left(\frac{1}{E-V}-\frac{2L^2_{}}{r^2_{}(E-V)^2_{}}\right)\nonumber \\ 
&+ & \left . V_{,r}\frac{3L^2_{}}{r^3_{}(E-V)}   \right\}\xi_G \; .\nonumber
\end{eqnarray}
\noindent The extra terms on the right have no physical meaning. The first thing to note is that the r.h.s. of eq.(\ref{Jcent}) vanishes for trajectories with  $\Delta=0$. Thus, we observe here again that the equation from the geometrical formalism with Jacobi metric is equivalent to equation from the tangent dynamics only for trajectories with constant kinetic energy. The reader might argue that although the equations are different, the stability analysis might still be equivalent. To show that this is not the case, it is enough to evaluate eq.(\ref{Jcent}) for $2\omega t=(2n+1)\frac{\pi}{2}$, to obtain
\begin{equation}
\ddot{\xi}_G+ \omega^2_{}\left(4-7\frac{\Delta^4_{}}{R^4_{}}\right)\xi_G=0\;, \quad \text{for}\; \; t=(2n+1)\frac{\pi}{2}.
\end{equation} 
 
\noindent The effective restoring force is negative in time intervals containing  $2\omega t=(2n+1)\frac{\pi}{2}$, for any trajectory with $\frac{\Delta^4_{}}{R^4_{}}>\frac{4}{7}$. The length of these ``explosive'' intervals is larger for larger values of $\Delta$.  Then, one finds again an infinite number of stable trajectories for which the geometrical approach with Jacobi metrics predicts an unstable behavior. This example shows that not only the effective frequencies become time dependent, they take imaginary values, periodically. 

In summary, the computation of the Lyapunov exponent within the kinetic energy metric has multiple shortcomings. In addition to the resulting equations being much more involved than those corresponding to the tangent dynamics, small kinetic energies make these equations numerically unstable. Moreover, the method by definition, is restricted to conservative systems. We also find that the sign of the curvature, and in general, the sign of any element of the curvature tensor, is not a good measure of stability. Lastly, the geodesic spread generally fails to describe the stability of the underlying physical system, by introducing un-physical parametric resonance and negative restoring forces.

\section{Conclusions}\label{conclus}
In this paper, we have shown that the vector field of geodesic spread ${\bm \xi}^{}_G$, and the tangent dynamic vector field ${\bm \xi}^{}_T$ are  equivalent when the arc length measured along any geodesic is proportional to the time interval. This is the case with the Eisenhart metric. When this is not fulfilled, as in the Jacobi metric, the geodesic spread is ill-defined. In the Jacobi metric, the  equations of motion satisfied by ${\bm \xi}^{}_G$  contain extra terms, that do not appear in ${\bm \xi}^{}_T$. These terms do not seem to have a clear  physical meaning, and can be responsible for the non-physical parametric resonance seen. Furthermore, they cause unstable modes (imaginary frequencies) in stable systems.

The Lyapunov exponent calculated within the geometrical formalism with Jacobi metric is equivalent to the tangent dynamics {\it if  the total kinetic energy is conserved}, which is unrealistic for interacting systems. The time evolution of the geodesic spread is not compatible with the movement in a constant-energy hypersurface.

  Using two representative examples, we demonstrated that the  geometrical Lyapunov exponent,  calculated with Jacobi metric, is correlated with the fluctuation of the kinetic energy, irrespective of the actual dynamical stability of the system. This implies that, in a statistical system of $N$ interacting particles, where the fluctuations in the kinetic energy decrease as $N$ increases, the formalism based on Jacobi metric would falsely predict that the system becomes ``less'' chaotic as $N\rightarrow \infty$. This, we believe, is the reason why an unexpected reduction of chaos with the increase of $N$ was seen in simulations  of a self-gravitating  system \cite{Pettini2}.

 We conclude that great care must be taken when using the Jacobi metric to derive stability results, especially in computing the Lyapunov exponents. In this respect the Eisenhart metric, or equivalently, the tangent dynamics remain as a simpler method that do not suffer from singularities.  It would be interesting to study whether other non-affine parameterizations have  similar drawbacks.

\end{document}